\documentclass[conference]{IEEEtran}
\IEEEoverridecommandlockouts
\usepackage{cite}
\usepackage{amsmath,amssymb,amsfonts}
\usepackage{graphicx}
\graphicspath{{figures/}}
\usepackage{textcomp}
\usepackage{xcolor}
\usepackage{booktabs}
\usepackage{enumitem}
\usepackage[hyphens]{url}
\usepackage[hidelinks]{hyperref}
\usepackage{xspace}
\usepackage{tikz}
\usetikzlibrary{arrows.meta,fit,positioning}
\usepackage{listings}
\def\BibTeX{{\rm B\kern-.05em{\sc i\kern-.025em b}\kern-.08em
    T\kern-.1667em\lower.7ex\hbox{E}\kern-.125emX}}

\newcommand{\sys}{NCCL M2N\xspace}

\begin{document}

\title{NCCL M2N: A Layout- and Topology-Aware\\
Collective for Distributed Tensor Resharding}

\author{\IEEEauthorblockN{Kaushik Kandadi, Youngeun Kwon, Sreeram Potluri, Ching-Hsiang Chu,\\
Ke Wen, Pouya Kousha, Sangkug Lym, Nitin Nitin, Manjunath Gorentla Venkata}
\IEEEauthorblockA{NVIDIA Corporation\\
\textit{\{kkandadisure, youngeunk, spotluri, chichu, kwen, pkousha, slym, nitin, manjunath\}@nvidia.com}}
}

\maketitle
\pagestyle{plain}

\begin{abstract}
Distributed deep learning training partitions tensors across a process mesh
according to a parallelism configuration, producing a corresponding sharding
and replication \emph{layout}. Reinforcement learning with verifiable rewards
(RLVR) often uses different layouts for policy training and rollout generation;
crossing this boundary requires \emph{resharding}
from one layout to another: the \emph{M-to-N layout redistribution} problem,
in which $M$ source ranks collectively hold a global tensor in one layout
and $N$ destination ranks need it in another. This pattern is absent from
today's collective libraries: flat direct sends duplicate every source
contribution across destination replicas, while gather-then-broadcast
concentrates traffic at one root and over-delivers the full tensor. These
workarounds are costly: with legacy all-gather plus broadcast, RL weight sync
accounts for 29.4\% of the reported RL step time for DeepSeek-V3 at 256 GPUs.

We present \sys{}, a collective primitive implemented in NCCL Extensions
for cross-layout sharded-tensor transfer. It automatically \emph{derives the
transfer schedule from the layout endpoints}. Compared with flat direct sends
and gather-then-broadcast, the collective provides: \textbf{(1) layout-derived traffic} --- only bytes
required by the destination layout are transferred; \textbf{(2)
load-balanced NIC usage} --- source contributions are distributed across
eligible destination leaders; and \textbf{(3) topology-aware routing} ---
replicated fan-out stays within an NVLink domain while a pipelined ring carries
one copy between domains.

We evaluate \sys{} on up to 256 GB200 GPUs in an NVL72 cluster with NDR
InfiniBand. On a single FFN-MoE layer transfer, \sys{} achieves up to
$\mathbf{7.9\times}$ speedup over flat direct sends (9.8\,ms vs.\ 77.3\,ms),
closely tracking the aggregate data-movement bound. In a non-colocated
DeepSeek-V3 NeMo-RL run, \sys{} reduces the reported weight-sync
time from 5.78\,s to 2.77\,s ($\mathbf{2.09\times}$), reducing step time by 12.7\%
over legacy all-gather plus broadcast.
\end{abstract}


\section{Introduction}

Distributed deep learning training partitions model tensors across graphics
processing units (GPUs) using various parallelism strategies such as data
parallelism (DP), tensor parallelism (TP), pipeline parallelism (PP), and
expert parallelism (EP)~\cite{megatron,megatron-pp,switch} to leverage the
high-performance computing power of GPU clusters.
As a result, these choices define each tensor's \emph{layout}: how the data
are sharded or replicated across the GPUs in the so-called process mesh.
Modern frameworks expose
this abstraction as \emph{distributed tensors}, including PyTorch
\textsc{DTensor}~\cite{dtensor}, GSPMD (general and scalable parallelization
for machine-learning computation graphs)~\cite{gspmd}, and Alpa~\cite{alpa}.
Application code can then reason about a tensor's global shape while the
runtime manages local placement.

Modern large language model (LLM) workflows routinely require \emph{resharding} tensors from one layout to
another as parallelism changes across phases. In such an M-to-N layout
reshard operation, $M$ source ranks collectively hold a tensor in one layout and
$N$ destination ranks require it in another. The runtime must derive the
overlapping tensor regions, deliver each region to every required replica, and
coordinate ranks that do not belong to the same process group. RLVR-style
training exposes this operation directly: it alternates rollout generation
and policy updates, often using different parallel layouts for the generator
and trainer. After each update, policy weights must be resharded and
transferred across the process-group boundary before the next rollout begins.
We refer to this synchronization phase as \emph{weight refit}. Recent
frameworks (NeMo-RL~\cite{nemo-rl}, OpenRLHF~\cite{openrlhf}, and
HybridFlow/verl~\cite{verl}) expose such a phase.
The RL weight refit step, which transfers updated policy weights from trainer
to generator after each training step, is the most costly
instance of this pattern. It becomes one of the major performance bottlenecks as
it sits directly on the critical path, where generators cannot begin the next rollout batch until new weights arrive.
Our profiling of DeepSeek-V3 at 256~GPUs showed that legacy all-gather plus
broadcast weight sync accounted for 29.4\% of the reported RL step time
(Table~\ref{tab:refit-cost}), a substantial share of the end-to-end step.

\begin{table}[t]
\centering
\caption{Legacy all-gather-plus-broadcast weight-sync cost as a fraction of
reported RL step time.}
\label{tab:refit-cost}
\small
\begin{tabular}{lrr}
\toprule
\textbf{Model} & \textbf{Scale} & \textbf{Refit share} \\
\midrule
DeepSeek-V3 & 256 GPUs & 29.4\% \\
\bottomrule
\end{tabular}
\end{table}

Despite this cost, today's communication stack does not express this
cross-group resharding operation natively.
The resulting workarounds expose two primary limitations:
\begin{enumerate}[leftmargin=*,label=(\roman*)]
    \item flat direct sends duplicate every source contribution across
    destination replicas; and
    \item gather-then-broadcast avoids source-side duplication but concentrates
    traffic at one root and over-delivers the full tensor, wasting network
    bandwidth.
\end{enumerate}

To address these problems, this paper proposes a layout-derived reshard collective primitive
and a topology-aware routing optimization. The collective derives one global transfer schedule from
the source and destination layouts, including exact overlap regions, replica
ownership, and disjoint process groups. The route sends one copy of each
contribution per destination NVLink domain, balances injection across eligible
NICs, forwards copies between domains, and performs replica fan-out locally
over NVLink. These choices remove replica-multiplied network traffic while
preserving the semantics of M-to-N layout changes expressible by the supported
    descriptors (\S\ref{sec:overview},~\S\ref{sec:algorithm}).

This paper makes the following contributions.
\begin{enumerate}[leftmargin=*,label=(\arabic*)]
    \item \textbf{A first-class M-to-N reshard collective primitive} with
    mesh-and-placement descriptors. It derives exact transfer regions across
    tensor dimensions, mismatched replica counts, cross-dimension sharding,
    and disjoint trainer/rollout process groups
    (Sections~\S\ref{sec:overview},~\S\ref{sec:api}).
    \item \textbf{A topology-aware routing optimization} that removes inter-node replica
    redundancy, balances NIC injection under shard imbalance, and performs
    local fan-out over NVLink while forwarding one copy across domains
    (Section~\S\ref{sec:algorithm}).
    \item \textbf{An evaluation of the hierarchical route} on an NVIDIA GB200
    NVL72 cluster. The measured route achieved up to a $\mathbf{7.9\times}$
    speedup for a single FFN-MoE layer transfer (9.8\,ms versus 77.3\,ms).
    In a separate non-colocated 256-GPU DeepSeek-V3 NeMo-RL run, \sys{}
    reduced the reported weight-sync time by $\mathbf{2.09\times}$
    and reduced step time by 12.7\% over legacy all-gather plus broadcast
    (Section~\S\ref{sec:eval}).
\end{enumerate}

\section{Background}
\label{sec:bg}

\subsection{RL Training Loop and Refit}
\label{sec:bg:rl-loop}

Reinforcement-learning fine-tuning alternates between rollout generation,
policy training, and weight synchronization. The policy uses a parallel
configuration optimized for training throughput and a separate generation
configuration optimized for rollout throughput. At each synchronization point, updated weights
move from the training layout to the generation layout. We call this in-memory
transfer phase \emph{refit}; it is an M-to-N layout redistribution~\cite{rfc-mxn}.

Refit combines different tensor layouts with communication between potentially
disjoint process groups and with destination-side replication. Because it runs
between training and the next rollout, its transfer schedule affects the
iteration critical path. The following subsections describe how frameworks
express these layouts and why existing communication abstractions do not
directly realize the transfer.

\subsection{Distributed Tensor Layouts}
\label{sec:bg:layouts}

Modern distributed-tensor frameworks can describe a global tensor with a
\emph{process mesh} and a per-tensor \emph{placement}. A placement maps each
mesh axis either to a tensor dimension (sharding) or to replication. PyTorch's
\textsc{DTensor}~\cite{dtensor}, GSPMD~\cite{gspmd}, and Alpa~\cite{alpa}
follow this model. For the materialized parameter tensors considered here, we
use \textsc{Shard} and \textsc{Replicate} placements. Given a global shape
$\mathbf{S}$, a mesh $M$, and a
placement list $p$ whose entries are \textsc{Shard}$(d)$ or
\textsc{Replicate}, each rank's local shape and offset are determined
deterministically. We call the tuple $(\mathbf{S}, M, p)$ a
\emph{layout}.

A \emph{reshard} converts a tensor from a source layout
$(\mathbf{S}, M_\text{src}, p_\text{src})$ to a destination layout
$(\mathbf{S}, M_\text{dst}, p_\text{dst})$. When the source and destination
groups contain $M$ and $N$ ranks, respectively, we call this an
\emph{M-to-N layout redistribution}. The two groups can use different shard
dimensions, shard counts, and replication factors, and can be disjoint. The
runtime must therefore derive which source regions overlap each destination
region instead of relying on a fixed participant pattern.

Tensor Parallelism (TP) and Expert Parallelism (EP) shard tensor and expert
dimensions, respectively. Pipeline Parallelism (PP) assigns successive layers
to different stages, while Data Parallelism (DP) replicates a model or tensor.
Combining these strategies gives training and generation layouts with
different ownership and replication patterns. In particular, TP and EP may
shard different dimensions, so a reshard can require intersecting
multi-dimensional regions rather than copying one rank's contiguous slice.

\subsection{Collective Communication and Interconnect Hierarchy}
\label{sec:bg:nccl}
\label{sec:bg:topology}

GPU-side distributed training relies on the NVIDIA Collective Communications
Library (NCCL)~\cite{nccl}, which provides \textsc{AllReduce},
\textsc{AllGather}, \textsc{ReduceScatter}, \textsc{Broadcast},
\textsc{AllToAll}, and point-to-point \texttt{send}/\texttt{recv}. A
collective API fixes its participants and logical input-to-output pattern;
the implementation can still choose topology-aware paths for that pattern.
For example, \textsc{AllGather} and \textsc{AllToAll} cover common in-mesh
transformations for materialized layouts when membership and sharded dimensions
align. \textsc{ReduceScatter} addresses a different transformation that
combines partial values with reduction; that \textsc{Partial} placement is
outside the layout subset used here. These collectives do not accept two
arbitrary layouts or derive the byte-level intersections needed when the source
and destination groups differ. No standard collective therefore expresses the
full cross-group M-to-N redistribution considered here.

NCCL also exposes lower-level mechanisms that can realize a custom route.
\emph{LSA} enables direct peer-memory loads and stores over supported peer
links, including within an NVLink domain.
\emph{GPU-Initiated Networking (GIN)}~\cite{gin} provides
GPU-initiated puts and signals over one-sided Remote Direct Memory Access
(RDMA), while host-initiated one-sided operations support staged transfers.
These mechanisms move data and signals, but they do not determine which layout
regions to send or where to create replicas; those decisions belong to the
reshard schedule. Because intra-domain links are typically faster than
cross-domain links, a topology-aware route should perform replica fan-out
within the destination domain and send each required tensor region once into
each destination domain.

\section{Motivation and Design Considerations}
\label{sec:overview}

\subsection{Limitations of Existing Approaches}
\label{sec:overview:motivation}
\label{sec:bg:motivation-primitives}

Cross-layout transfers are often assembled from existing lower-level
communication operations, with two primary limitations:
\begin{enumerate}[leftmargin=*,label=(\roman*)]
    \item flat direct sends duplicate every source contribution across
    destination replicas; and
    \item gather-then-broadcast avoids source-side duplication but concentrates
    traffic at one root and over-delivers the full tensor, wasting network
    bandwidth.
\end{enumerate}
The missing capability is therefore not another send operation. The runtime
needs one collective that jointly plans the M-to-N
redistribution from both endpoint layouts, identifies the overlapping tensor
slices, maps each slice to source and destination byte offsets, and accounts for
the communication topology.

\subsection{Layout-Derived Schedule and Hierarchical Route}
\label{sec:overview:approach}

We define a \emph{layout-derived schedule} as a transfer plan computed at call
time from the source and destination layout descriptors together with the
runtime topology. The descriptors determine which tensor regions overlap and
which destination replicas require each region. The topology determines which
leaders and links carry those regions.

For the layout class formalized in Section~\ref{sec:api}, the hierarchical
route provides three properties.
First, it removes replica-multiplied source egress by injecting each required
contribution once and forwarding one copy through each destination domain.
Second, it balances the number of source contributions assigned to each
eligible leader rather than concentrating them on one receiver. Third, it keeps
replica forwarding within an NVLink domain and uses inter-domain links only for
the copies that must cross them.
These properties eliminate replica-multiplied network traffic and
single-ingress over-delivery while enabling global topology-aware load
balancing and preserving the semantics of the supported M-to-N layouts
(Section~\ref{sec:api}).

The next sections formalize the layout and overlap semantics, then derive the
schedule and hierarchical route that realize these properties.

\section{Reshard Semantics and Descriptors}
\label{sec:api}

\tikzset{
  rk/.style={draw, rounded corners=2pt, fill=white,
    minimum width=0.76cm, minimum height=0.38cm,
    font=\scriptsize, inner sep=1pt, line width=0.55pt},
  ldr/.style={rk, line width=1.4pt, fill=blue!5},
  sil/.style={rk, draw=gray!45, dashed, fill=gray!9, text=gray!55},
  rdma/.style={-{Stealth[length=3pt]}, line width=0.85pt,
               draw=red!75!black},
  rdup/.style={-{Stealth[length=3pt]}, line width=0.75pt,
               draw=red!55!white, dashed},
  nvlink/.style={-{Stealth[length=3pt]}, line width=0.9pt,
                 draw=blue!60},
}

To dive deeper into the semantics of resharding, we first introduce the layout model
and relevant terminology as shown in Table~\ref{tab:layout-model-summary}.

\paragraph*{Process mesh}
A \emph{process mesh} $M$ is a multidimensional grid of GPU ranks with a
deterministic mapping from mesh coordinates to communicator ranks. For the
two-dimensional $D_0 \times D_1$ case used in our examples, indices $(i_0,\, i_1)$
select a slot with $0 \le i_k < D_k$. The grid is anchored at a
\emph{base rank} $r_0$ within the communicator used for this reshard; slots
form a contiguous row-major rank interval,
\begin{equation}
  \mathrm{rank}(i_0,\, i_1) \;=\; r_0 \;+\; i_0\, D_1 \;+\; i_1 \, .
\label{eq:mesh-rank}
\end{equation}

\paragraph*{Placements and layouts}
A \emph{placement} list $p$ fixes how each active mesh axis relates to
tensor storage; a \emph{layout} $L = (\mathbf{S},\, M,\, p)$ pairs that
mesh with a global shape $\mathbf{S} = (S_0, \ldots, S_{n-1})$.
Table~\ref{tab:layout-model-summary} summarizes the layout notation.

\begin{table}[t]
\centering
\caption{Notation for the layout model.}
\label{tab:layout-model-summary}
\footnotesize
\setlength{\tabcolsep}{4pt}
\begin{tabular}{@{}p{0.34\columnwidth}p{0.60\columnwidth}@{}}
\toprule
\textbf{Symbol} & \textbf{Meaning} \\
\midrule
$L = (\mathbf{S},\, M,\, p)$
  & Complete layout comprising a global tensor shape, process mesh, and
    placement list. \\
$\mathbf{S} = (S_0,\ldots,S_{n-1})$
  & Global tensor shape; $S_d$ is the extent of tensor dimension $d$. \\
$M,\; D_k,\; r_0$
  & Process mesh; $D_k$ is the number of ranks along mesh axis $k$, and
    $r_0$ is the mesh's base rank within the communicator. \\
$i_k$
  & Coordinate along mesh axis $k$, with $0 \le i_k < D_k$. \\
$p_k = \textsc{Shard}(d)$
  & Axis $k$ partitions tensor dimension $d$ into $D_k$ contiguous
    pieces; each rank stores a $1/D_k$ slice along~$d$. \\
$p_k = \textsc{Replicate}$
  & Axis $k$ duplicates data: every rank along that axis holds the same
    slice; it does not subdivide any tensor dimension. \\
\bottomrule
\end{tabular}
\end{table}

\paragraph*{Local shard bounds}
To determine which portion of the global tensor a rank owns, consider
the rank at mesh coordinates $\mathbf{i}$. Its \emph{local shard} spans
$[l_d,\, u_d)$ along each tensor dimension~$d$. The lower endpoint is
\begingroup
\footnotesize
\begin{equation}
  l_d \;=\;
  \begin{cases}
    \bigl\lfloor i_k \cdot S_d \;/\; D_k \bigr\rfloor
      & \exists k:\, p_k = \textsc{Shard}(d), \\
    0 & \text{otherwise (\textsc{Replicate}-only).}
  \end{cases}
\label{eq:local-lo}
\end{equation}
\endgroup
The upper endpoint $u_d$ uses the same formula with $(i_k{+}1)$ in place
of $i_k$. Supported sharded dimensions divide evenly by their shard count;
replicated dimensions use the full extent $[0,\, S_d)$. The public API
accepts each side's local shape rather than $\mathbf{S}$ and verifies that
the two descriptors imply the same global shape.

\paragraph*{Reshard}
A \emph{reshard} from $L_s = (\mathbf{S},\, M_s,\, p_s)$ to
$L_d = (\mathbf{S},\, M_d,\, p_d)$ is a collective that, when complete,
leaves every destination rank $r \in M_d$ holding the local shard of
$\mathbf{S}$ prescribed by~$L_d$.
The implementation derives the full (source rank, destination rank,
byte-range) plan from the two endpoints; callers never supply a manual
per-pair schedule. Every communicator rank supplies both endpoint layout
descriptions, $L_s$ and $L_d$, so that all ranks derive the same schedule.
For each endpoint, a rank supplies a pointer to its local tensor buffer only
if it participates in that endpoint's mesh; otherwise, only the layout
metadata is required. The abstract layout model does not require disjoint
meshes. The current implementation, however, requires $M_s$ and $M_d$ to be
disjoint intervals within the supplied communicator.

Next, we use two running examples to illustrate the semantics of resharding.
Example~A highlights destination-side replication and same-dimension TP
consolidation: a TP=4 trainer (DP=1) refitting weights to a TP=2,
DP=2 generator.
Example~B shows how pipeline parallelism (PP$\,{>}\,1$) is handled via
per-stage decomposition: a TP\,=\,2, PP\,=\,2 trainer refitting weights to
a TP\,=\,1, PP\,=\,1 generator.

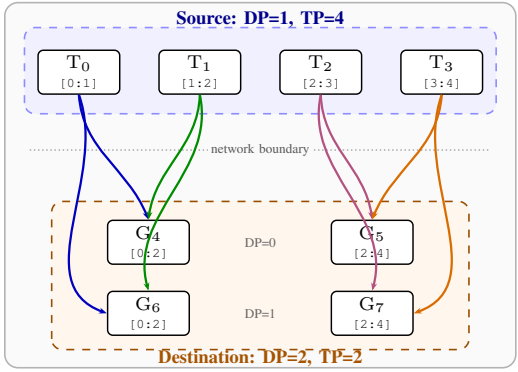
\begin{figure}[t]
\centering
\begin{tikzpicture}[
  flowrk/.style={rk, minimum width=1.08cm, minimum height=0.58cm,
                 align=center},
  net0/.style={-{Stealth[length=3pt]}, line width=0.85pt,
               draw=blue!75!black},
  net1/.style={-{Stealth[length=3pt]}, line width=0.85pt,
               draw=green!55!black},
  net2/.style={-{Stealth[length=3pt]}, line width=0.85pt,
               draw=magenta!70!black},
  net3/.style={-{Stealth[length=3pt]}, line width=0.85pt,
               draw=orange!85!black}
]

\fill[gray!4, rounded corners=5pt]
    (-3.38,-1.48) rectangle (3.38,3.35);
\draw[black!30, rounded corners=5pt, line width=0.6pt]
    (-3.38,-1.48) rectangle (3.38,3.35);
\fill[blue!6, rounded corners=4pt]
    (-3.12,1.88) rectangle (3.12,3.02);
\draw[blue!45, dashed, rounded corners=4pt, line width=0.6pt]
    (-3.12,1.88) rectangle (3.12,3.02);
\node[font=\scriptsize\bfseries, color=blue!55!black]
    at (0,3.13) {Source: DP=1, TP=4};
\fill[orange!7, rounded corners=4pt]
    (-2.76,-1.24) rectangle (2.76,0.72);
\draw[orange!60!black, dashed, rounded corners=4pt, line width=0.6pt]
    (-2.76,-1.24) rectangle (2.76,0.72);
\node[font=\scriptsize\bfseries, color=orange!65!black]
    at (0,-1.35) {Destination: DP=2, TP=2};
\draw[black!35, densely dotted, line width=0.55pt]
    (-3.05,1.40) -- (3.05,1.40);
\node[font=\tiny, fill=gray!4, inner sep=1.2pt, text=black!65]
    at (0,1.40) {network boundary};

\node[flowrk] (AT0) at (-2.40,2.43)
    {$\mathrm{T}_0$\\[-1pt]{\tiny\texttt{[0:1]}}};
\node[flowrk] (AT1) at (-0.80,2.43)
    {$\mathrm{T}_1$\\[-1pt]{\tiny\texttt{[1:2]}}};
\node[flowrk] (AT2) at ( 0.80,2.43)
    {$\mathrm{T}_2$\\[-1pt]{\tiny\texttt{[2:3]}}};
\node[flowrk] (AT3) at ( 2.40,2.43)
    {$\mathrm{T}_3$\\[-1pt]{\tiny\texttt{[3:4]}}};

\node[flowrk] (AG4) at (-1.48,0.18)
    {$\mathrm{G}_4$\\[-1pt]{\tiny\texttt{[0:2]}}};
\node[flowrk] (AG5) at ( 1.48,0.18)
    {$\mathrm{G}_5$\\[-1pt]{\tiny\texttt{[2:4]}}};
\node[flowrk] (AG6) at (-1.48,-0.76)
    {$\mathrm{G}_6$\\[-1pt]{\tiny\texttt{[0:2]}}};
\node[flowrk] (AG7) at ( 1.48,-0.76)
    {$\mathrm{G}_7$\\[-1pt]{\tiny\texttt{[2:4]}}};
\node[font=\tiny, text=black!60] at (0,0.18) {DP=0};
\node[font=\tiny, text=black!60] at (0,-0.76) {DP=1};

\draw[net0] (AT0.south) to[out=-86,in=112] (AG4.north);
\draw[net0] (AT0.south) to[out=-70,in=160] (AG6.west);
\draw[net1] (AT1.south) to[out=-92,in=82] (AG4.north);
\draw[net1] (AT1.south) to[out=-78,in=105] (AG6.north);
\draw[net2] (AT2.south) to[out=-88,in=98] (AG5.north);
\draw[net2] (AT2.south) to[out=-102,in=75] (AG7.north);
\draw[net3] (AT3.south) to[out=-94,in=68] (AG5.north);
\draw[net3] (AT3.south) to[out=-110,in=20] (AG7.east);

\end{tikzpicture}
\caption{Flat direct-send realization of Example~A's required ownership. Each source
contribution is sent to both destination DP replicas, so the route transfers
$2B$ over the network. Colors distinguish the four source contributions.}
\label{fig:exa-mesh}
\end{figure}

\paragraph*{Example~A: same-dimension TP reshard with destination replicas.}
Let $\mathbf{S} = [4,\, 8]$.
The source mesh places four ranks $\{0\text{--}3\}$ in a $[1{\times}4]$
grid ($r_0^s = 0$) with
$p_s = \bigl(\textsc{Replicate},\; \textsc{Shard}(0)\bigr)$:
axis-0 replicates trivially (DP=1) and axis-1 shards dimension~0 into
four equal slices (TP=4), so each source rank holds one row of~$\mathbf{S}$.
The destination mesh places four ranks $\{4\text{--}7\}$ in a $[2{\times}2]$
grid ($r_0^d = 4$) with
$p_d = \bigl(\textsc{Replicate},\; \textsc{Shard}(0)\bigr)$:
axis-0 replicates (DP=2) and axis-1 shards dimension~0 into two slices
(TP=2).
Table~\ref{tab:example-a} and Figure~\ref{fig:exa-mesh} show the
resulting shard ownership and mesh layout.

\begin{table}[t]
\centering
\caption{Shard ownership for Example~A ($\mathbf{S}\!=\![4,8]$,
mesh $[1{\times}4]\!\to\![2{\times}2]$): source and destination both
shard dimension~0; each destination TP slot consolidates two source
TP shards and is replicated twice (DP=2).
\texttt{[a:b,\;c:d]}: dim-0/dim-1 slice of $\mathbf{S}$.}
\label{tab:example-a}
\footnotesize
\setlength{\tabcolsep}{4pt}
\begin{tabular}{@{}cclp{0.32\columnwidth}@{}}
\toprule
\textbf{Rank} & \textbf{Role} & \textbf{Local shard} & \textbf{Parallelism} \\
\midrule
0 & src & \texttt{[0:1,\; 0:8]} & TP=0 \\
1 & src & \texttt{[1:2,\; 0:8]} & TP=1 \\
2 & src & \texttt{[2:3,\; 0:8]} & TP=2 \\
3 & src & \texttt{[3:4,\; 0:8]} & TP=3 \\
\midrule
4 & dst & \texttt{[0:2,\; 0:8]} & TP=0, DP=0 \\
5 & dst & \texttt{[2:4,\; 0:8]} & TP=1, DP=0 \\
6 & dst & \texttt{[0:2,\; 0:8]} & TP=0, DP=1 (replica of rank~4) \\
7 & dst & \texttt{[2:4,\; 0:8]} & TP=1, DP=1 (replica of rank~5) \\
\bottomrule
\end{tabular}
\end{table}

Since both meshes shard along dimension~0, each destination TP slot
consolidates two adjacent source TP shards:
$\mathrm{G}_4$ receives rows $[0\text{:}1]$ and $[1\text{:}2]$ from
$\mathrm{T}_0$ and $\mathrm{T}_1$;
$\mathrm{G}_5$ receives rows $[2\text{:}3]$ and $[3\text{:}4]$ from
$\mathrm{T}_2$ and $\mathrm{T}_3$.
Destination row~1 ($\mathrm{G}_6$, $\mathrm{G}_7$) holds data identical
to row~0 (generator-side DP=2 replication): $\mathrm{G}_6$ duplicates
$\mathrm{G}_4$ and $\mathrm{G}_7$ duplicates $\mathrm{G}_5$.
Figure~\ref{fig:exa-mesh} shows a flat direct-send realization of this ownership:
each source contribution is sent to both DP copies. This transfers two copies
of the unique payload over the network and motivates the topology-aware route
described in \S\ref{sec:alg:leaders}.

\paragraph*{Example~B: PP$\,{>}\,1$ via per-stage reshard decomposition.}

With pipeline parallelism (PP$\,{>}\,1$), model layers are partitioned
across PP stages; each stage's ranks own weights for only their assigned
layers.
Because different stages hold entirely different weight tensors, no single
reshard can span the full trainer mesh---instead, the caller issues
\emph{one reshard per PP stage}, each scoped to that stage's ranks plus
the generator, leaving the reshard primitive itself unchanged.

Consider four trainer ranks (TP\,=\,2, PP\,=\,2, DP\,=\,1) refitting to
a single generator rank (TP\,=\,1, PP\,=\,1).
For a representative weight tensor $\mathbf{S} = [4,\, 8]$
(same shape as Example~A for direct comparison):
PP stage~0 assigns ranks~$\{0,\,1\}$ in a $[1{\times}2]$ mesh with
$p_s = (\textsc{Replicate},\;\textsc{Shard}(0))$ (TP\,=\,2, no DP);
PP stage~1 assigns ranks~$\{2,\,3\}$ to an identical mesh arrangement.
The generator (rank~4) uses a $[1{\times}1]$ mesh with
$p_d = (\textsc{Replicate})$ and must hold the full tensor for every
stage's layers.
Table~\ref{tab:example-pp} lists the shard ownership for both reshard calls.
After both calls complete, $\mathrm{G}_4$ holds two separate full tensors of
shape $[4,8]$---one for each PP stage's layers---whereas each trainer rank
carries only the TP shard for its own stage.
No five-rank communicator is required for this decomposition. Each reshard
operates within a dedicated \emph{PP communicator} scoped to
its participants: PP comm~0 on global ranks~$\{0,\,1,\,4\}$ and PP comm~1
on~$\{2,\,3,\,4\}$, with local comm ranks $0,\,1,\,2$ within each.
The mesh's $r_0$ (Eq.~\ref{eq:mesh-rank}) is therefore a rank within its
PP communicator---rank~4 appears as comm rank~2 in both.
Every rank in a given PP communicator still calls \texttt{ncclReshard} and
passes both descriptors, as required by the collective contract. Because
there is no destination-side replication (DP\,=\,1 on the generator),
each call reduces to a pure TP consolidation---two shards merging into one
full tensor---structurally identical to a single-stage version of Example~A.

\begin{table}[t]
\centering
\caption{Shard ownership for Example~B ($\mathbf{S}\!=\![4,8]$,
trainer TP\,=\,2 PP\,=\,2, generator TP\,=\,1 PP\,=\,1).
Each reshard uses its own PP communicator; \textbf{c-rank} is the
rank's local index within that communicator.
Rank~4 (global) joins both PP comms as c-rank~2, accumulating one
full-shape tensor per stage (s0/s1).}
\label{tab:example-pp}
\footnotesize
\setlength{\tabcolsep}{3pt}
\begin{tabular}{@{}cccclc@{}}
\toprule
\textbf{Rank} & \textbf{Role} & \textbf{PP comm} & \textbf{c-rank} & \textbf{Local shard} & \textbf{TP slot} \\
\midrule
\multicolumn{6}{@{}l@{}}{\itshape Reshard call~1\;---\;PP comm~0 $\{$ranks 0,\,1,\,4$\}$} \\[1pt]
0 & src & 0 & 0 & \texttt{[0:2,\; 0:8]} & TP=0 \\
1 & src & 0 & 1 & \texttt{[2:4,\; 0:8]} & TP=1 \\
4 & dst & 0 & 2 & \texttt{[0:4,\; 0:8]} & full (s0) \\
\midrule
\multicolumn{6}{@{}l@{}}{\itshape Reshard call~2\;---\;PP comm~1 $\{$ranks 2,\,3,\,4$\}$} \\[1pt]
2 & src & 1 & 0 & \texttt{[0:2,\; 0:8]} & TP=0 \\
3 & src & 1 & 1 & \texttt{[2:4,\; 0:8]} & TP=1 \\
4 & dst & 1 & 2 & \texttt{[0:4,\; 0:8]} & full (s1) \\
\bottomrule
\end{tabular}
\end{table}

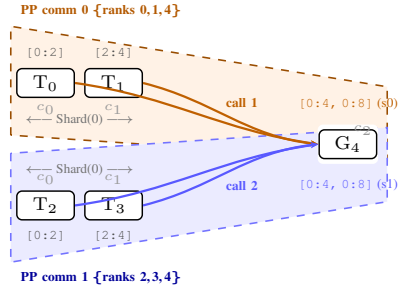
\begin{figure}[t]
\centering
\begin{tikzpicture}

\fill[orange!10]
    (-0.46, 0.12) -- (-0.46, 1.56) -- (4.52, 0.76) -- (4.52, -0.22) -- cycle;
\draw[orange!60!black, dashed, line width=0.6pt]
    (-0.46, 0.12) -- (-0.46, 1.56) -- (4.52, 0.76) -- (4.52, -0.22) -- cycle;
\node[font=\tiny\bfseries, color=orange!70!black, anchor=south west]
    at (-0.45, 1.56) {PP comm~0\;\{ranks 0,\,1,\,4\}};

\fill[blue!8]
    (-0.46, -0.12) -- (-0.46, -1.56) -- (4.52, -0.76) -- (4.52, 0.22) -- cycle;
\draw[blue!55, dashed, line width=0.6pt]
    (-0.46, -0.12) -- (-0.46, -1.56) -- (4.52, -0.76) -- (4.52, 0.22) -- cycle;
\node[font=\tiny\bfseries, color=blue!60!black, anchor=north west]
    at (-0.45, -1.56) {PP comm~1\;\{ranks 2,\,3,\,4\}};

\fill[white, rounded corners=2pt] (3.60, -0.26) rectangle (4.40, 0.26);

\node[rk] (T0) at (0.00, 0.81) {\scriptsize $\mathrm{T}_0$};
\node[rk] (T1) at (0.90, 0.81) {\scriptsize $\mathrm{T}_1$};
\node[rk] (T2) at (0.00,-0.81) {\scriptsize $\mathrm{T}_2$};
\node[rk] (T3) at (0.90,-0.81) {\scriptsize $\mathrm{T}_3$};

\node[font=\tiny, color=black!70] at (0.00, 1.21) {\texttt{[0:2]}};
\node[font=\tiny, color=black!70] at (0.90, 1.21) {\texttt{[2:4]}};
\node[font=\tiny, color=black!70] at (0.00,-1.21) {\texttt{[0:2]}};
\node[font=\tiny, color=black!70] at (0.90,-1.21) {\texttt{[2:4]}};

\node[font=\tiny, color=black!40] at (0.00, 0.46) {$c_0$};
\node[font=\tiny, color=black!40] at (0.90, 0.46) {$c_1$};
\node[font=\tiny, color=black!40] at (0.00,-0.46) {$c_0$};
\node[font=\tiny, color=black!40] at (0.90,-0.46) {$c_1$};

\node[font=\tiny, color=black!55]
    at (0.45, 0.33) {$\longleftarrow$\,Shard(0)\,$\longrightarrow$};
\node[font=\tiny, color=black!55]
    at (0.45,-0.33) {$\longleftarrow$\,Shard(0)\,$\longrightarrow$};

\node[rk] (G4) at (4.00, 0.00) {\scriptsize $\mathrm{G}_4$};
\node[font=\tiny, color=black!40] at (4.20, 0.19) {$c_2$};
\node[font=\tiny, color=orange!75!black] at (4.00, 0.54)
    {\texttt{[0:4,\;0:8]}\;(s0)};
\node[font=\tiny, color=blue!65] at (4.00,-0.54)
    {\texttt{[0:4,\;0:8]}\;(s1)};

\draw[-{Stealth[length=3pt]}, line width=0.85pt, draw=orange!75!black]
    (T0.east) to[out=-8,  in=172] (G4.west);
\draw[-{Stealth[length=3pt]}, line width=0.85pt, draw=orange!75!black]
    (T1.east) to[out=-14, in=174] (G4.west);

\draw[-{Stealth[length=3pt]}, line width=0.85pt, draw=blue!65]
    (T2.east) to[out=8,   in=-172] (G4.west);
\draw[-{Stealth[length=3pt]}, line width=0.85pt, draw=blue!65]
    (T3.east) to[out=14,  in=-174] (G4.west);

\node[font=\tiny\bfseries, color=orange!75!black] at (2.60, 0.55)
    {call~1};
\node[font=\tiny\bfseries, color=blue!65] at (2.60,-0.55)
    {call~2};

\end{tikzpicture}
\caption{PP comm structure for Example~B (trainer TP\,=\,2 PP\,=\,2,
generator TP\,=\,1 PP\,=\,1, global ranks~0--4).
Each reshard uses its own \emph{PP communicator}; a communicator containing
all trainer stages is unnecessary.
Orange funnel: PP comm~0 on global ranks~$\{0,\,1,\,4\}$ (c-ranks
$0,\,1,\,2$); blue funnel: PP comm~1 on~$\{2,\,3,\,4\}$ (same local
numbering).
$\mathrm{G}_4$ sits at the overlap---it is c-rank~2 in both---and
accumulates one full tensor per stage (s0 via call~1, s1 via call~2).
The mesh's $r_0$ (Eq.~\ref{eq:mesh-rank}) is a rank within its PP
communicator, not the global rank space.}
\label{fig:pp-mesh}
\end{figure}

Figure~\ref{fig:pp-mesh} shows the mesh decomposition.
Both PP comm groups share rank~4 as their common destination; the
generator mesh is constructed once and reused across both calls.

Listing~\ref{lst:pp-reshard} makes each rank's communicator membership
explicit. Each trainer invokes \texttt{nccl.m2n.reshard} once, while generator
rank~4 invokes it twice. Both stages share the same communicator-local layout
in this example; callers may supply stage-specific layouts.

\begin{lstlisting}[language=Python,float=t,captionpos=b,
abovecaptionskip=8pt,belowcaptionskip=6pt,
caption={Packaged Python API for PP decomposition. Each PP communicator uses
local ranks 0--1 for its source mesh and local rank~2 for its destination mesh.},
label={lst:pp-reshard}]
import nccl.m2n as m2n

# PP communicators joined by this rank:
# ranks 0-1: [(0, pp_comm0)]
# ranks 2-3: [(1, pp_comm1)]
# rank 4:    [(0, pp_comm0), (1, pp_comm1)]
for stage, pp_comm in local_stage_comms:
    src_mesh = m2n.Mesh([2], start_rank=0)
    dst_mesh = m2n.Mesh([1], start_rank=2)
    src_tensor = local_stage_weights[stage] if is_trainer else None
    dst_tensor = generator_weights[stage] if is_generator else None
    m2n.reshard(
        src_tensor, dst_tensor, pp_comm, stream=stream,
        src_mesh=src_mesh,
        src_placements=[m2n.Shard(0)],
        dst_mesh=dst_mesh,
        dst_placements=[m2n.Replicate()],
    )
\end{lstlisting}

\section{Layout-Derived Design and Implementation}
\label{sec:algorithm}

\sys{} separates \emph{what} bytes move from \emph{how} they move. From the
endpoint layouts, the planner derives the required byte ranges and
topology-aware route; the implementation realizes that schedule using the
mechanisms described at the end of this section.

\subsection{Deterministic Schedule Derivation}
\label{sec:alg:algorithm}

Section~\ref{sec:api} defines each rank's interval $[l_\delta,u_\delta)$
along tensor dimension $\delta$. For source $s$ and destination $t$, write
these bounds as $[l_\delta^s,u_\delta^s)$ and $[l_\delta^t,u_\delta^t)$.
Their overlap is
\[
I_\delta(s,t)=
[\max(l_\delta^s,l_\delta^t),\,\min(u_\delta^s,u_\delta^t)).
\]
For each source rank $s$, the planner tests every destination rank $t$. If every
$I_\delta(s,t)$ is nonempty, the combined intervals define the exact
contribution; otherwise, the pair is skipped. Replicated source copies collapse
to one deterministic owner. This common enumeration gives all ranks the same
schedule without coordination.

\subsection{Hierarchical Topology Route}
\label{sec:alg:leaders}

The schedule above maps each source rank to overlapping destination shards and,
conversely, each destination shard to its unique source contributions. Flat
direct sends transmit every contribution separately to all replicas of its destination
shard. The hierarchical route instead balances contributions across selected
replica leaders: each contribution enters through one leader, is ring-forwarded
across domains, and is replicated locally.

Among $K$ destination shards, let $j$ identify one, $q_j$ its number of unique
source contributions, and $\rho$ its local replicas per domain. Ordered by
source shard, each replica receives base load
$a_j=\lfloor q_j/\rho\rfloor$; the first $b_j=q_j\bmod\rho$ receive one extra.
Replicas with nonzero loads become leaders and forward their contributions to
the other $\rho-1$ replicas, so loads differ by at most one.

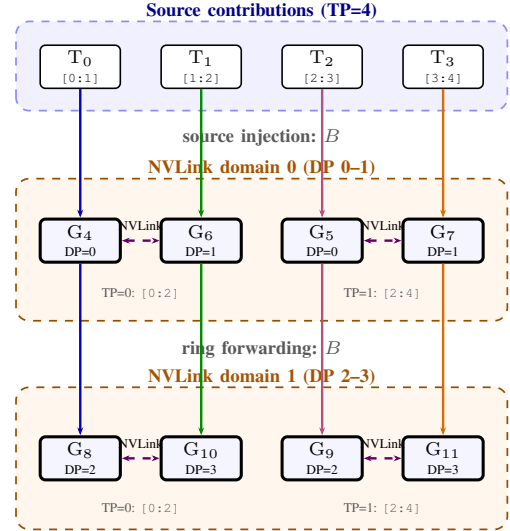
\begin{figure}[t]
\centering
\begin{tikzpicture}[
  topork/.style={draw, rounded corners=2pt, fill=white,
                 minimum width=1.06cm, minimum height=0.57cm,
                 font=\scriptsize, align=center, inner sep=1.5pt,
                 line width=0.6pt},
  leader/.style={topork, line width=1.15pt, fill=blue!4},
  net0/.style={-{Stealth[length=3pt]}, line width=0.9pt,
               draw=blue!75!black},
  net1/.style={-{Stealth[length=3pt]}, line width=0.9pt,
               draw=green!55!black},
  net2/.style={-{Stealth[length=3pt]}, line width=0.9pt,
               draw=magenta!70!black},
  net3/.style={-{Stealth[length=3pt]}, line width=0.9pt,
               draw=orange!85!black},
  localx/.style={{Stealth[length=3pt]}-{Stealth[length=3pt]},
                 dashed, line width=0.9pt, draw=violet!80!black}
]

\fill[blue!6, rounded corners=5pt]
    (-3.25,4.18) rectangle (3.25,5.36);
\draw[blue!45, dashed, rounded corners=5pt, line width=0.65pt]
    (-3.25,4.18) rectangle (3.25,5.36);
\node[font=\scriptsize\bfseries, color=blue!55!black]
    at (0,5.49) {Source contributions (TP=4)};

\fill[orange!7, rounded corners=5pt]
    (-3.25,1.40) rectangle (3.25,3.28);
\draw[orange!60!black, dashed, rounded corners=5pt, line width=0.65pt]
    (-3.25,1.40) rectangle (3.25,3.28);
\node[font=\scriptsize\bfseries, color=orange!65!black]
    at (0,3.42) {NVLink domain 0 (DP 0--1)};

\fill[orange!7, rounded corners=5pt]
    (-3.25,-1.36) rectangle (3.25,0.52);
\draw[orange!60!black, dashed, rounded corners=5pt, line width=0.65pt]
    (-3.25,-1.36) rectangle (3.25,0.52);
\node[font=\scriptsize\bfseries, color=orange!65!black]
    at (0,0.66) {NVLink domain 1 (DP 2--3)};

\node[topork] (HT0) at (-2.40,4.76)
    {$\mathrm{T}_0$\\[-1pt]{\tiny\texttt{[0:1]}}};
\node[topork] (HT1) at (-0.80,4.76)
    {$\mathrm{T}_1$\\[-1pt]{\tiny\texttt{[1:2]}}};
\node[topork] (HT2) at (0.80,4.76)
    {$\mathrm{T}_2$\\[-1pt]{\tiny\texttt{[2:3]}}};
\node[topork] (HT3) at (2.40,4.76)
    {$\mathrm{T}_3$\\[-1pt]{\tiny\texttt{[3:4]}}};

\node[leader] (HG4) at (-2.40,2.46) {$\mathrm{G}_4$\\[-1pt]{\tiny DP=0}};
\node[leader] (HG6) at (-0.80,2.46) {$\mathrm{G}_6$\\[-1pt]{\tiny DP=1}};
\node[leader] (HG5) at (0.80,2.46) {$\mathrm{G}_5$\\[-1pt]{\tiny DP=0}};
\node[leader] (HG7) at (2.40,2.46) {$\mathrm{G}_7$\\[-1pt]{\tiny DP=1}};
\node[font=\tiny, text=black!65] at (-1.60,1.77)
    {TP=0: \texttt{[0:2]}};
\node[font=\tiny, text=black!65] at (1.60,1.77)
    {TP=1: \texttt{[2:4]}};

\node[leader] (HG8)  at (-2.40,-0.42) {$\mathrm{G}_8$\\[-1pt]{\tiny DP=2}};
\node[leader] (HG10) at (-0.80,-0.42) {$\mathrm{G}_{10}$\\[-1pt]{\tiny DP=3}};
\node[leader] (HG9)  at (0.80,-0.42) {$\mathrm{G}_9$\\[-1pt]{\tiny DP=2}};
\node[leader] (HG11) at (2.40,-0.42) {$\mathrm{G}_{11}$\\[-1pt]{\tiny DP=3}};
\node[font=\tiny, text=black!65] at (-1.60,-1.08)
    {TP=0: \texttt{[0:2]}};
\node[font=\tiny, text=black!65] at (1.60,-1.08)
    {TP=1: \texttt{[2:4]}};

\node[font=\scriptsize\bfseries, fill=white, inner sep=1pt,
      text=black!65] at (0,3.82) {source injection: $B$};
\draw[net0] (HT0.south) -- (HG4.north);
\draw[net1] (HT1.south) -- (HG6.north);
\draw[net2] (HT2.south) -- (HG5.north);
\draw[net3] (HT3.south) -- (HG7.north);

\node[font=\scriptsize\bfseries, fill=white, inner sep=1pt,
      text=black!65] at (0,1.02) {ring forwarding: $B$};
\draw[net0] (HG4.south) -- (HG8.north);
\draw[net1] (HG6.south) -- (HG10.north);
\draw[net2] (HG5.south) -- (HG9.north);
\draw[net3] (HG7.south) -- (HG11.north);

\draw[localx] (HG4.east) -- node[above, font=\tiny] {NVLink} (HG6.west);
\draw[localx] (HG5.east) -- node[above, font=\tiny] {NVLink} (HG7.west);
\draw[localx] (HG8.east) -- node[above, font=\tiny] {NVLink} (HG10.west);
\draw[localx] (HG9.east) -- node[above, font=\tiny] {NVLink} (HG11.west);

\end{tikzpicture}
\caption{Representative hierarchical route for DP=4 across two destination
NVLink domains ($d=2$, $\rho=2$). Solid arrows show $B$-byte source injection
and ring forwarding; dashed bidirectional arrows show local NVLink exchange
reconstructing each TP slice on both replicas.}
\label{fig:hier-route}
\end{figure}

Figure~\ref{fig:hier-route} extends Example~A to DP=4 across two domains,
where $K=2$, $\rho=2$, and $q_0=q_1=2$, so $a_0=a_1=1$ and
$b_0=b_1=0$. Thus $\mathrm{T}_0$/$\mathrm{T}_1$ enter through
$\mathrm{G}_4$/$\mathrm{G}_6$ and $\mathrm{T}_2$/$\mathrm{T}_3$ through
$\mathrm{G}_5$/$\mathrm{G}_7$, balancing all four NICs rather than
concentrating receives on DP0's $\mathrm{G}_4$ and $\mathrm{G}_5$. Each
network stage carries $B$; chunking is omitted, and traffic scales with
destination domains, not replicas.

\subsection{Aggregate Data-Movement Model}
\label{sec:alg:bw-analysis}

The following model gives a data-movement lower bound for the route evaluated
in \S\ref{sec:eval}. It intentionally omits local data preparation, chunking,
protocol, and startup costs, so the bound is not a prediction of every
component of end-to-end latency.

Let $B$ be the unique global payload, $N_s$ the number of active source GPUs,
$K$ the number of distinct destination shards, and $r_d$ the destination
replica count. The destination spans $d$ NVLink domains, with
$\rho=r_d/d$ local replicas per shard in each domain. Let $\beta_N$ and
$\beta_V$ be the usable one-way network and NVLink bandwidth per GPU.

\paragraph*{Hierarchical route.}
The source injects one copy of $B$. A destination domain has $K\rho$ active
receivers, so the aggregate bandwidth of the network stage is
\begin{equation}
\mathcal{B}_N
=\min\!\left(N_s\beta_N,\;K\rho\beta_N\right),
\qquad
T_\text{network}=\frac{B}{\mathcal{B}_N}.
\label{eq:network-time}
\end{equation}
When the destination spans multiple domains, the inter-domain ring forwards
that copy through successive domains. The network stages operate concurrently
for a large payload, so $d$ affects startup and tail latency but does not
multiply the steady-state $B/\mathcal{B}_N$ term.

Within each destination domain, the network supplies one copy and NVLink
creates the remaining $\rho-1$ copies. The local copy volume is therefore
$(\rho-1)B$. Fair leader selection lets all $K\rho$ destination GPUs
participate, giving
\begin{equation}
T_\text{NVLink}
=\frac{(\rho-1)B}{K\rho\beta_V}.
\label{eq:nvlink-time}
\end{equation}
Network movement and local replication overlap, hence the data-movement
speed-of-light (SOL) bound is simply
\begin{equation}
\boxed{T_\text{SOL}
=\max\!\left(T_\text{network},T_\text{NVLink}\right).}
\label{eq:hier-time}
\end{equation}

\paragraph*{Flat direct sends.}
Flat direct sends inject the payload once per destination replica. Their
aggregate network bound and transfer time are
\begin{equation}
\mathcal{B}_N^\text{flat}
=\min\!\left(N_s\beta_N,\;Kr_d\beta_N\right),
\qquad
T_\text{flat}=\frac{r_dB}{\mathcal{B}_N^\text{flat}}.
\label{eq:flat-time}
\end{equation}
When source injection is the bottleneck and NVLink remains below the network
term, dividing Eqs.~\eqref{eq:flat-time} and~\eqref{eq:hier-time} gives
\begin{equation}
\boxed{\frac{T_\text{flat}}{T_\text{SOL}}\approx r_d.}
\label{eq:speedup}
\end{equation}

\paragraph*{All-gather plus broadcast.}
A common framework path first all-gathers the source shards so that a designated
trainer root holds the full payload, then broadcasts that payload to the
destination ranks. These stages are typically pipelined across bounded buffers.
Ignoring pipeline fill and drain, and assuming the two stages do not contend for
resources, the ideal steady-state bound is
\begin{equation}
T_\text{AG+Bcast}^\text{SOL}
\approx \max\!\left(T_\text{AG}^\text{SOL},
                    T_\text{Bcast}^\text{SOL}\right).
\label{eq:ag-bcast-time}
\end{equation}
When the broadcast crosses the inter-mesh network and root injection is the
bottleneck, $T_\text{Bcast}^\text{SOL}=B/\beta_N$. In the source-limited regime,
M2N injects the $N_s$ source shards concurrently, giving
$T_\text{M2N}^\text{SOL}=B/(N_s\beta_N)$. Therefore,
\begin{equation}
\boxed{T_\text{Bcast}^\text{SOL}
       \approx N_s T_\text{M2N}^\text{SOL}.}
\label{eq:ag-bcast-ratio}
\end{equation}
Here $N_s$ is the number of independent source-injection paths; it equals the
model-shard count when every shard can inject independently, and should be
replaced by the effective path count when shards share NIC bandwidth.

Section~\ref{sec:eval:single} instantiates these aggregate expressions and
checks explicitly that NVLink is not on the critical path.

\subsection{Implementation and Availability}
\label{sec:impl}

The initial release (v0.2) supports one- and two-dimensional process meshes,
one- to three-dimensional tensors, and at most one sharded mesh axis per
layout. It realizes the derived schedule through two hierarchical transports. PACK
gathers strided contributions into contiguous staging buffers, executes the
hierarchical transfer, and scatters them into destination buffers. PIPE,
currently beta, pipelines bounded chunks through the same route using
device-side or host-driven remote-memory-access (RMA) progress.
Both transports preserve caller-stream ordering. \sys{} is released as the
\texttt{nccl\_m2n} component of the open-source NVIDIA NCCL Extensions
repository.\footnote{\url{https://github.com/NVIDIA/nccl-extensions}}
It provides C and packaged Python interfaces for layout-derived resharding over
ordinary CUDA buffers; the repository documents the current API, build,
configuration, and transport details.

\section{Evaluation}
\label{sec:eval}

\subsection{Experimental Setup}
\label{sec:eval:setup}

Experiments use a GB200 NVL72 cluster with NDR InfiniBand
(50\,GB/s per NIC per direction). The single FFN-MoE layer experiment uses NCCL
Extensions' \texttt{reshard\_model\_bench} to isolate a 7168\,MB BF16 expert
gate-projection transfer. A 16-GPU EP trainer group occupies one NVLink domain,
while a TP8 generator group with DP$\in\{2,4,8\}$ (16--64 GPUs) occupies a
separate domain. This expert-sharded-to-tensor-parallel transfer represents a
dominant MoE weight-movement pattern. The end-to-end
experiment uses DeepSeek-V3 in BF16 on 256 GPUs across 64 nodes of the same
cluster. The 128 trainer GPUs span two NVL72 domains, and the 128 non-colocated
generator GPUs span two additional domains, with 64 active GPUs per domain.
Trainer parallelism is TP=1, PP=8, EP=16; generator parallelism is TP=16, DP=8.
Single-layer latency is the maximum across ranks of each rank's average over
10 timed iterations following three warmups; every point also passes three
correctness-validation iterations. The flat direct-send baseline sends from
each active source rank to every final destination replica independently.

\paragraph*{Measurement provenance.}
The single FFN-MoE layer measurements were rerun with NCCL 2.31.0-1 and the NCCL
Extensions model benchmark. They compare the flat direct-send baseline with
the hierarchical route. The end-to-end experiment compares \sys{} against
NeMo-RL's legacy all-gather-plus-broadcast path under the same
non-colocated asynchronous RL configuration. It uses one-step off-policy
execution and sequence length 1536. End-to-end metrics are reported for
non-warmup execution and are labeled as step time and weight-sync time.

\subsection{Single-Layer Microbenchmark}
\label{sec:eval:single}

\begin{figure}[t]
\centering
\includegraphics[width=\columnwidth]{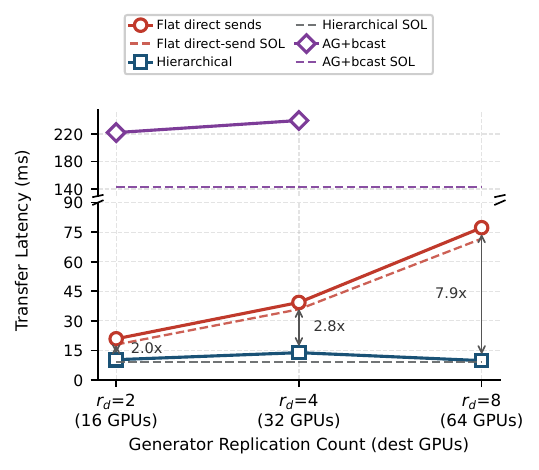}
\caption{Single FFN-MoE layer transfer latency on GB200 NVL72 + NDR.
Trainer: TP=1, EP=16, DP=$\{1,2,4\}$, with one node-aware active EP
replica; generator: TP=8, DP=$r_d$ (16--64 GPUs in a separate NVLink
domain). Tensor: $[256, 2048, 7168]$ BF16, 7168\,MB total. Dashed lines
show the analytical SOL model (Eqs.~\eqref{eq:hier-time},
\eqref{eq:flat-time}, and~\eqref{eq:ag-bcast-time}); markers and solid lines
show measured results. The broken y-axis separates the AG+bcast values from
the lower-latency paths. AG+bcast measurements are included for $r_d{=}2$ and
$r_d{=}4$; the $r_d{=}8$ point is SOL-only. Arrows annotate hierarchical
speedup over flat direct sends. The
Hierarchical series reports the lowest validated hierarchical-transport result
at each point.}
\label{fig:single-layer}
\end{figure}

Figure~\ref{fig:single-layer} sweeps the generator replication count
$r_d \in \{2, 4, 8\}$ (corresponding to 16--64 destination GPUs) while
holding the active source set fixed at $m\!=\!16$ EP shards and the tensor
parallelism at $K\!=\!8$. Additional trainer replicas make the two sides
equal in allocation size but remain inactive under node-aware assignment.

\paragraph*{Model bandwidth validation.}
We instantiate the aggregate equations with $B{=}7168$\,MB,
$N_s{=}16$, $K{=}8$, and $\beta_N{=}50$\,GB/s. All three points fit in
one destination NVLink domain, so the corresponding local replica counts are
$\rho{=}\{2,4,8\}$. In every case,
\begin{align*}
\mathcal{B}_N
  &=\min(16\!\times\!50,\;8\rho\!\times\!50)
    =800\;\text{GB/s}, \\
T_\text{network}
  &=\frac{7168}{800}=8.96\;\text{ms}.
\end{align*}

A GB200 GPU provides 1.8\,TB/s bidirectional NVLink bandwidth, for which we
use the one-way value $\beta_V{=}900$\,GB/s~\cite{gb200-nvl72}. Equation
\eqref{eq:nvlink-time} then gives
\begin{equation*}
T_\text{NVLink}
=\frac{(\rho-1)7168}{8\rho\!\times\!900}
=\{0.50,\;0.75,\;0.87\}\;\text{ms}.
\end{equation*}
Thus NVLink is below the network term even at eight-way local replication,
and Eq.~\eqref{eq:hier-time} gives $T_\text{SOL}=8.96$\,ms for all three
points. No per-chunk timing is needed for this bound.

Flat direct sends instead inject $r_dB$ through the same source aggregate
bandwidth:
\begin{equation*}
T_\text{flat}
  =\frac{r_d\!\times\!7168}{800}
  =8.96r_d\;\text{ms},
\end{equation*}
\begin{equation*}
T_\text{flat}
  =\{17.9,\;35.8,\;71.7\}\;\text{ms}.
\end{equation*}
The predicted ratio $T_\text{flat}/T_\text{SOL}=r_d$ yields the dashed curves
in Figure~\ref{fig:single-layer}.

The rank-0 AG+bcast path first all-gathers the 16 EP shards within the trainer
NVLink domain, then broadcasts the full payload from one trainer root to all
generator ranks. Applying Eq.~\eqref{eq:ag-bcast-time}, the ideal 16-way ring
all-gather and root-injection bounds are
\begin{align*}
T_\text{AG}^\text{SOL}
  &\approx \frac{15}{16}\frac{7168}{900}=7.47\;\text{ms}, \\
T_\text{Bcast}^\text{SOL}
  &=\frac{7168}{50}=143.36\;\text{ms}.
\end{align*}
Assuming full pipeline overlap and no resource contention, broadcast dominates,
so $T_\text{AG+Bcast}^\text{SOL}=143.36$\,ms for all three replication counts.
This is $N_s{=}16$ times the 8.96\,ms M2N SOL, matching
Eq.~\eqref{eq:ag-bcast-ratio}.

\paragraph*{Hierarchical transport latency is nearly constant.}
The best validated hierarchical path completes the three transfers in
$\{10.251,13.834,9.775\}$\,ms despite the $4\times$ increase in generator
replication. These values are respectively 14\%, 54\%, and 9\% above the
8.96\,ms SOL. The residual includes contiguous staging, pack/unpack,
signalling, and startup costs. The nearly flat series confirms that adding
destination replicas does not proportionally load the source NIC.

\paragraph*{Flat direct-send latency grows with $r_d$.}
Flat direct-send latency rises from 20.859\,ms at $r_d\!=\!2$ to 77.258\,ms at
$r_d\!=\!8$ --- a $3.7\times$ increase as replication grows $4\times$.
The three values are only 16\%, 10\%, and 8\% above the corresponding
$\{17.92,35.84,71.68\}$\,ms flat direct-send SOL values. This decreasing
relative excess is consistent with a roughly fixed startup cost becoming a
smaller fraction of the growing flat direct-send transfer volume.

\paragraph*{Rank-0 AG+bcast is root-injection limited.}
For $r_d\!=\!2$ and 4, the validated rank-0 packed path completes in
222.065 and 239.649\,ms, or $1.55\times$ and $1.67\times$ the 143.36\,ms SOL.
The corresponding effective root-injection bandwidths are 32.3 and
29.9\,GB/s (64.6\% and 59.8\% of the assumed 50\,GB/s bandwidth). The
implementation uses two bounded world broadcasts totaling 7.00\,GiB;
splitting the payload does not change the steady-state byte bound. The
$r_d\!=\!8$ measurement is pending, so Figure~\ref{fig:single-layer} shows only
its analytical AG+bcast SOL value.

\paragraph*{Speedup tracks the aggregate SOL bound.}
Hierarchical speedup over flat direct sends grows from $2.0\times$ at
$r_d\!=\!2$ to
$7.9\times$ at $r_d\!=\!8$, with $2.8\times$ at $r_d\!=\!4$. The endpoint
values closely approach the analytical $\{2,8\}\times$ prediction of
Eq.~\eqref{eq:speedup}); the middle point retains more transport overhead.

\subsection{End-to-End DeepSeek-V3 RL Weight Sync}
\label{sec:eval:e2e}

\begin{figure}[t]
\centering
\includegraphics[width=\columnwidth]{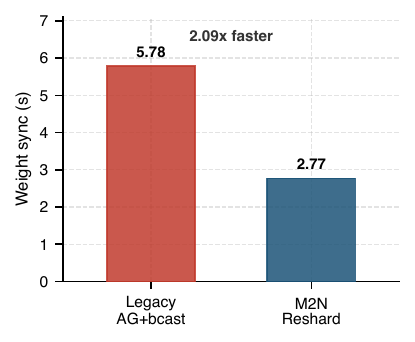}
\caption{DeepSeek-V3 NeMo-RL weight-sync time on 256 GB200 GPUs.
Trainer: TP=1, PP=8, EP=16 (128 GPUs); generator: TP=16, DP=8
(128 GPUs). Each group spans two separate NVL72 domains, with 64 active GPUs
per domain. Measurements use BF16, sequence length 1536, and one-step
off-policy asynchronous execution.}
\label{fig:e2e-refit}
\end{figure}

Figure~\ref{fig:e2e-refit} compares the reported weight-sync time for
DeepSeek-V3 in BF16 using asynchronous NeMo-RL. Both runs use the topology
described in \S\ref{sec:eval:setup}, sequence length 1536, one-step
off-policy execution, and non-colocated trainer and generator groups.
The comparison is between \sys{} and legacy all-gather plus broadcast.

\paragraph*{Weight-sync latency.}
M2N reduces the reported weight-sync time from 5.78\,s to 2.77\,s, a
52.1\% reduction or $\mathbf{2.09\times}$ speedup. In the baseline,
weight sync accounts for 29.4\% of the reported 19.68\,s step time,
providing the motivation in Table~\ref{tab:refit-cost}.

\paragraph*{Step time and stability.}
The reported step time decreases from 19.68\,s to 17.18\,s (12.7\%).
Weight-sync time falls by 3.01\,s, while step time falls by 2.50\,s;
the two metrics need not change by the same amount in asynchronous execution.
The reported KL error is 0.001869 with M2N and 0.001912 with the legacy path.
These close values provide a limited stability check for this comparison,
rather than evidence of convergence or training quality.

\section{Related Work}
\label{sec:related}

\paragraph*{Distributed-tensor abstractions and RL training systems.}
PyTorch DTensor~\cite{dtensor}, GSPMD~\cite{gspmd}, and
Alpa~\cite{alpa} make layout a first-class abstraction but provide no
cross-mesh collective.
\texttt{torchstore}~\cite{torchstore} and PyTorch
Monarch~\cite{monarch} both bridge meshes through DTensor-aware
transfer interfaces. torchstore additionally provides topology-based
transport selection through a put/get interface. \sys{} addresses a different
layer by deriving one global schedule from both endpoint layouts and
coordinating all shard groups jointly.
TensorHub~\cite{tensorhub} layers a storage abstraction
(Reference-Oriented Storage) above pair-wise transport, tracking
where weights live and serving topology-optimized reads on demand;
\sys{} shares this goal of topology-awareness but operates at a
different layer --- a single collective call with a global,
layout-derived schedule rather than a directory of per-replica
fetches.
NeMo-RL~\cite{nemo-rl}, OpenRLHF~\cite{openrlhf},
HybridFlow/veRL~\cite{verl}, and LlamaRL~\cite{llamarl} each
implement their own refit phase using a mix of NCCL broadcast,
CUDA IPC, and one-sided DMA --- efficient for their target setup,
but treating refit as opaque pair-wise sends rather than as a
planned topology-aware collective.
\sys{} provides the missing primitive in this stack: an NCCL-native
collective for cross-layout sharded-tensor transfer that derives
a single global schedule from two (mesh, placement) endpoints at
call time and plans jointly across the interconnect hierarchy ---
a drop-in replacement for refit and other cross-layout transfers
that preserves each framework's process-group structure.

\paragraph*{Programmable communication abstractions.}
MSCCL++~\cite{mscclpp} exposes channel and signal abstractions over
NVLink and RDMA; NVSHMEM~\cite{nvshmem} provides OpenSHMEM-style
one-sided primitives. The GPU-Initiated Networking interface for
NCCL~\cite{gin} unifies NVLink and RDMA paths in NCCL's device API
and is the foundation our data plane builds on. NCCLX~\cite{ncclx}
scales standard collectives to 100K+ GPUs without introducing
layout-aware primitives. \sys{} contributes the layer above these
abstractions: a layout-derived schedule that tells GIN/LSA/multimem
\emph{what} to move and \emph{which path} to use.

\paragraph*{Offline checkpoint resharding.}
Universal Checkpointing~\cite{ucp} saves a parallelism-agnostic
checkpoint and reconstructs any target layout on load;
ByteCheckpoint~\cite{bytecheckpoint} disaggregates data and metadata
for load-time resharding with $\sim$3.6$\times$ speedup over manual
scripts. Both target the file-system path across runs, while \sys{}
targets in-memory transfer between two live process groups within a
single run --- the two approaches are complementary.

\section{Discussion and Future Work}
\label{sec:discussion}

\sys{} establishes a deployable API and several transport choices, but
the following directions remain open.

\paragraph*{Multi-dimensional and higher-rank sharding.}
The current implementation permits at most one sharded mesh axis per layout.
Extending schedule derivation to support simultaneous
sharding across multiple dimensions --- and meshes with more than two
axes --- would cover arbitrary GSPMD-style placements and three-way
parallelism combinations (e.g., TP$\times$EP$\times$DP) natively,
without composing multiple two-dimensional meshes.

\paragraph*{Elastic communicator APIs.}
Production RL systems increasingly resize their rollout clusters
dynamically. Emerging NCCL extensions for \emph{shrink} and \emph{grow}
communicators allow process group membership to change without a full
teardown and reinitialize cycle. Integrating these APIs into \sys{}
would allow the reshard schedule to be updated in-place as the source
or destination mesh changes size, making \sys{} a natural weight-transfer
primitive for elastic RL training loops.

\paragraph*{Adaptive transport selection.}
Contiguous staging and device- or host-driven remote-memory access expose
different storage and startup costs.
A policy can choose among them
per transfer using contiguity, tensor size, peer count, staging pressure, and
whether several compatible calls can be grouped. Such a policy should retain
the current explicit environment override for reproducible evaluation.

\paragraph*{Overlapping meshes and recovery.}
Multi-rank source and destination meshes are currently disjoint, and protocol
errors after host-RMA entry require communicator-level shutdown. Supporting
overlapping membership would broaden in-place layout transitions; distributed
failure agreement would allow safe recovery instead of the current fail-stop
contract. Both require protocol changes beyond descriptor validation.

\paragraph*{On-the-fly quantization.}
Quantization is commonly applied to model weights before transfer to
reduce bandwidth consumption. Rather than quantizing weights as a
separate pass before resharding, \sys{}'s staging kernel can be extended
to apply quantization transformations in-flight as chunks move through
the staging ring --- fusing the two operations into a single kernel pass
and eliminating the intermediate quantized buffer.

\section{Conclusion}
\label{sec:conclusion}

We presented \sys{}, a collective primitive for transferring distributed
tensors between different source and destination layouts. By deriving the
transfer schedule from both layouts, it sends only the tensor regions required
by each destination and balances network traffic across destination replicas.
Its hierarchical route avoids the repeated network transfers of flat direct
sends and the full-tensor gathering of gather-then-broadcast, using NVLink to
complete replication within each destination domain.

On a GB200 NVL72 cluster, \sys{} achieved up to $7.9\times$ speedup over flat
direct sends for a single FFN-MoE layer transfer. In a separate 256-GPU
DeepSeek-V3 NeMo-RL experiment, it reduced weight-sync time from 5.78\,s to
2.77\,s---a $2.09\times$ speedup over legacy all-gather plus broadcast---and
reduced step time by 12.7\%. These results demonstrate the benefit of planning
tensor resharding jointly across layouts and the interconnect hierarchy.

\bibliographystyle{IEEEtran}
\bibliography{references}

\end{document}